\newcommand{\al}{\mbox{$^{26}$\hspace{-0.2em}Al}}
\newcommand{\Msol}{M_{\sun}}

\newcommand{\gray}{\mbox{$\gamma$-ray}}

\def\MeV{\mbox{Me\hspace{-0.1em}V}}

\def\deg{\hbox{$^\circ$}}
\def\sun{\hbox{$\odot$}}

\def\aj{AJ}

\def\aaps{A\&AS}

\def\procspie{Proc.~SPIE}

\def   \ni {\noindent}

\def   \ssk {\vskip  5truept}

\def   \bsk {\vskip 15truept}

\def   \newpage {\vfill\eject}
\def   \newline {\hfil\break}

\documentstyle[epsfig]{article}
\begin{document}
\hsize 5truein
\vsize 8truein
\font\abstract=cmr8
\font\keywords=cmr8
\font\caption=cmr8
\font\references=cmr8
\font\text=cmr10
\font\affiliation=cmssi10
\font\author=cmss10
\font\mc=cmss8
\font\title=cmssbx10 scaled\magstep2
\font\alcit=cmti7 scaled\magstephalf
\font\alcin=cmr6
\font\ita=cmti8
\font\mma=cmr8
\def\ref{\par\noindent\hangindent 15pt}
\null

\title{\ni A New Look at Image Reconstruction of COMPTEL 1.8 \MeV\ Data}

\bsk \bsk
\author{\ni J.~Kn\"odlseder$^1$,
            D.~Dixon$^6$,
            R.~Diehl$^2$,
            U.~Oberlack$^5$,
	        P.~von Ballmoos$^1$,
            H.~Bloemen$^3$, \\
            W.~Hermsen$^3$, 
            A.~Iyudin$^2$,
	        J.~Ryan$^4$,
	        and
	        V.~Sch\"onfelder$^2$}
\bsk
\affiliation{$^1$Centre d'Etude Spatiale des Rayonnements, CNRS/UPS, B.P.~4346,
	         31028 Toulouse Cedex 4, France}
\affiliation{$^2$Max-Planck-Institut f\"ur extraterrestrische Physik,
	         Postfach 1603, 85740 Garching, Germany}
\affiliation{$^3$SRON-Utrecht, Sorbonnelaan 2, 3584 CA Utrecht, The
	         Netherlands}
\affiliation{$^4$Space Science Center, University of New Hampshire, Durham
	         NH 03824, U.S.A.}
\affiliation{$^5$Astrophysics Laboratory, Columbia University, New York, 
	         NY 10027, U.S.A.}
\affiliation{$^6$Institute for Geophysics and Planetary Physics, 
	         University of California, Riverside, CA 92521, U.S.A.}
\bsk
\baselineskip = 12pt

\abstract{ABSTRACT \ni 
Deconvolving COMPTEL \gray\ data into images presents a major 
methodological challenge.
We developed a new algorithm called Multiresolution Regularized Expectation 
Maximization (MREM), which explicitly accounts for spatial 
correlations in the image by using wavelets (Dixon et al.~1998).
We demonstrate that MREM largely suppresses image noise in the 
reconstruction, showing only significant structures that are present 
in the data.
Application to 1.8 \MeV\ data results in a sky map that is much smoother 
than the maximum entropy reconstructions presented previously,
but it shows the same characteristic emission features which have been 
established earlier.
}
\bsk
\baselineskip = 12pt
\keywords{\ni KEYWORDS: image reconstruction;
                        gamma-ray lines; 
                        wavelets
}

\bsk
\baselineskip = 12pt

\text{
\ni 1. INTRODUCTION
\ssk
\ni
\label{sec:intro}

Imaging the \gray\ sky at MeV energies is one of the important 
milestones achieved by the COMPTEL telescope aboard the {\em Compton 
Gamma-Ray Observatory} ({\em CGRO}).
In particular, COMPTEL provided the first image of the sky in the light 
of the 1.809 \MeV\ \gray\ line radiation, attributed to radioactive decay 
of \al\ (Diehl et al.~1995; Oberlack et al.~1996).
Reconstruction of intensity distributions from the complexly encoded 
COMPTEL data requires a deconvolution procedure.
The sky maps presented in previous works were generally obtained 
using a maximum entropy (ME) algorithm (Strong et al.~1992).
Starting from a grey image, ME iteratively solves for a 
maximum entropy solution with increasingly structure entering the 
reconstruction as the iterations proceed.
Unfortunately, there is no objective criterion where to stop the 
iterations, and the resulting image depends on the choice of the 
data analyst.
Additionally, it has been shown that the reconstructions are very 
sensitive to statistical noise in the data, leading to lumpy images
even in the presence of a smooth and structureless \gray\ intensity 
distribution (Kn\"odlseder et al.~1996).
\linebreak \newpage \noindent

For these reasons, we developed a new algorithm called 
{\em Multiresolution Regularized Expectation Maximization}, which
explicitly accounts for spatial correlations in the reconstructed 
image by using wavelets.
This effectively reduces the number of free parameters in the 
reconstruction process which is at the origin of the lumpy aspect 
of ME images.
MREM is based on the Richardson-Lucy algorithm (Richardson 1972; Lucy 
1974) which is a special case of the more general class of expectation 
maximization (EM) algorithms (Dempster et al.~1977).
We introduced an additional wavelet thresholding step in the iterative 
EM scheme which aims to extract the significant structure in the 
reconstruction.
This leads to a convergent algorithm which automatically stops when the
significant structure has been extracted from the data.
For a detailed description and discussion of the algorithm the reader 
is referred to Dixon et al.~(1998).

\bsk
\ni 2. SIMULATIONS
\ssk
\ni
\label{sec:simulations}

To illustrate the performance of MREM with respect to ME we apply both 
algorithms to simulated COMPTEL observations of 
(1) an exponential disk distribution normalized to $3\Msol$ of total 
galactic \al\ mass, and
(2) the EGRET $>100$ MeV all-sky map adjusted to a plausible 1.809 \MeV\ 
intensity level by fitting to COMPTEL 1.8 \MeV\ data.
The first case tests the response to a smooth and structureless 
intensity distribution while the second case is probably a more 
realistic situation with some point sources embedded in a structured 
diffuse emission distribution.
The resulting all-sky maps and longitude profiles are shown in Fig.~1.
It is clearly visible that in both cases ME leads to lumpy 
reconstructions.
In particular, only few emission spots in the reconstruction of the 
EGRET sky map coincide indeed with real sources -- most of the spots 
are artefacts due to statistical noise, appearing arbitrarily along 
the galactic plane.
MREM drastically improves above ME in that it reasonably reconstructs 
the emission profiles without introducing artificial hot spots.
The latitude extent of the reconstructions is slightly higher than that 
of the models, but this is a combined result of the instrument's angular 
resolution of only $4\deg$ (FWHM) together with a low signal to noise 
ratio.
Yet, e.g.~the extended diffuse emission above the galactic center is 
still recovered in the EGRET MREM map.

\bsk
\ni 3. THE COMPTEL 1.8 \MeV\ SKY
\ssk
\ni
\label{sec:comptel}

\begin{figure}[h]
\centerline{\psfig{file=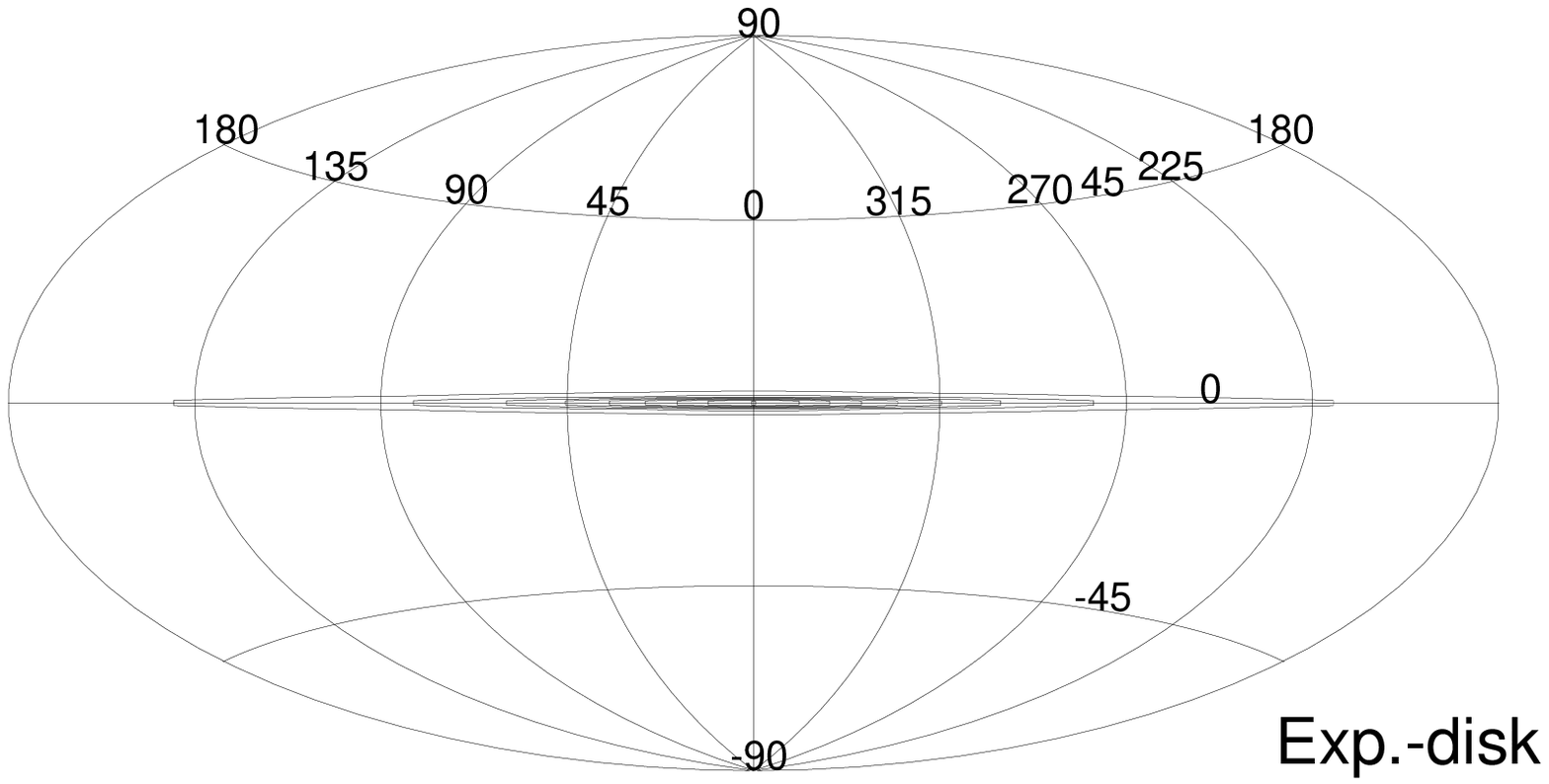, width=6.3cm, height=3cm}
            \hfill
            \psfig{file=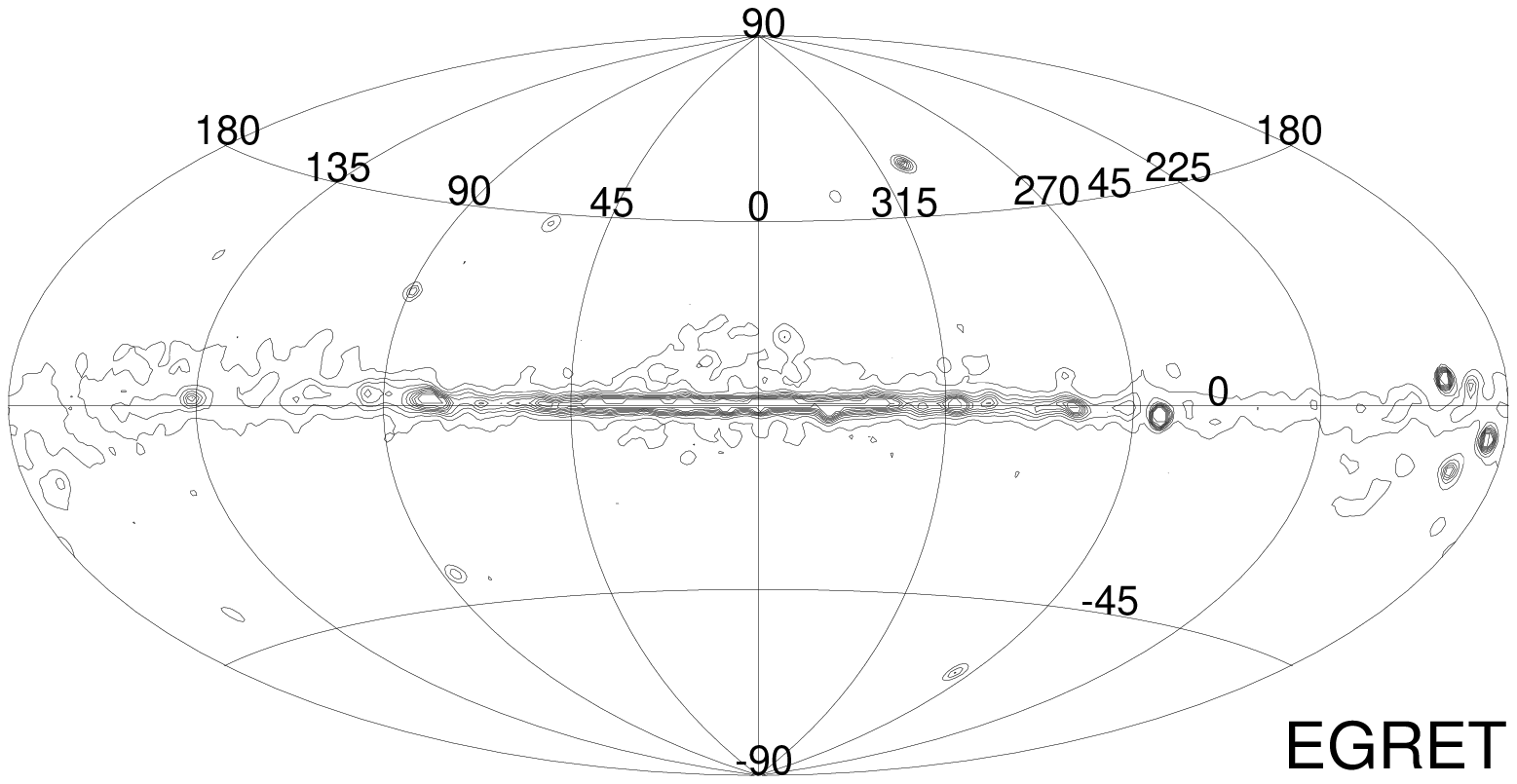, width=6.3cm, height=3cm}}
\centerline{\psfig{file=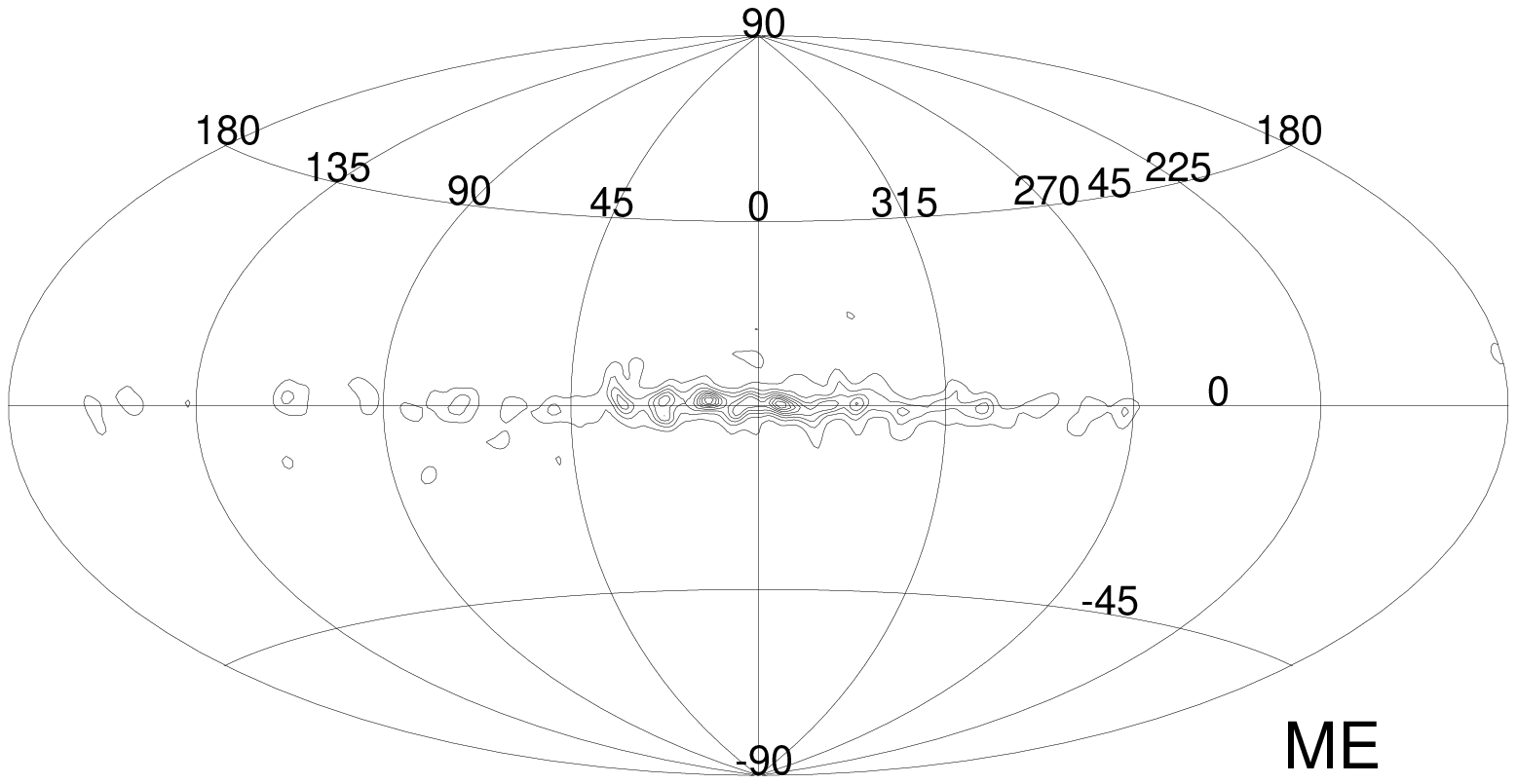, width=6.3cm, height=3cm}
            \hfill
            \psfig{file=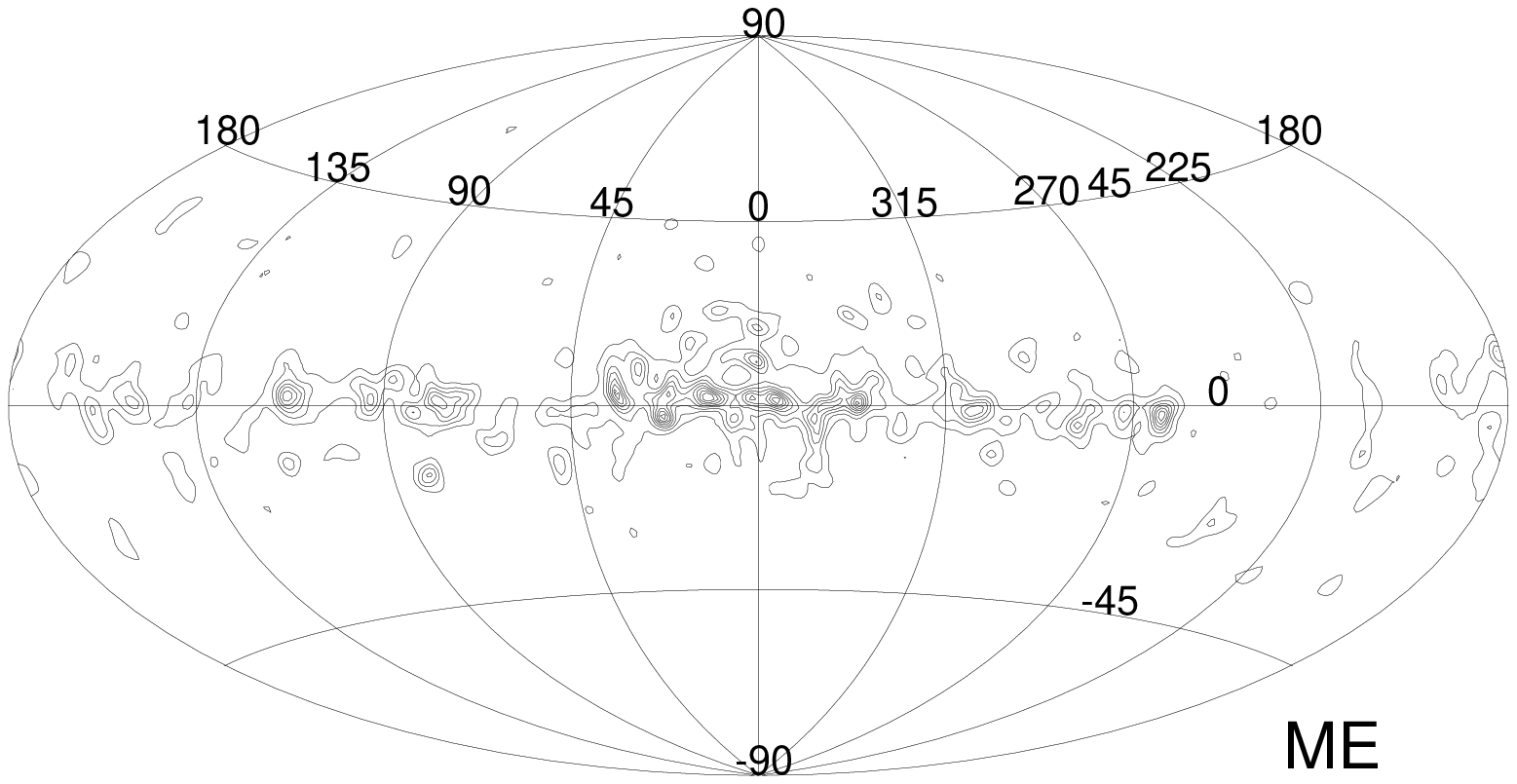, width=6.3cm, height=3cm}}
\centerline{\psfig{file=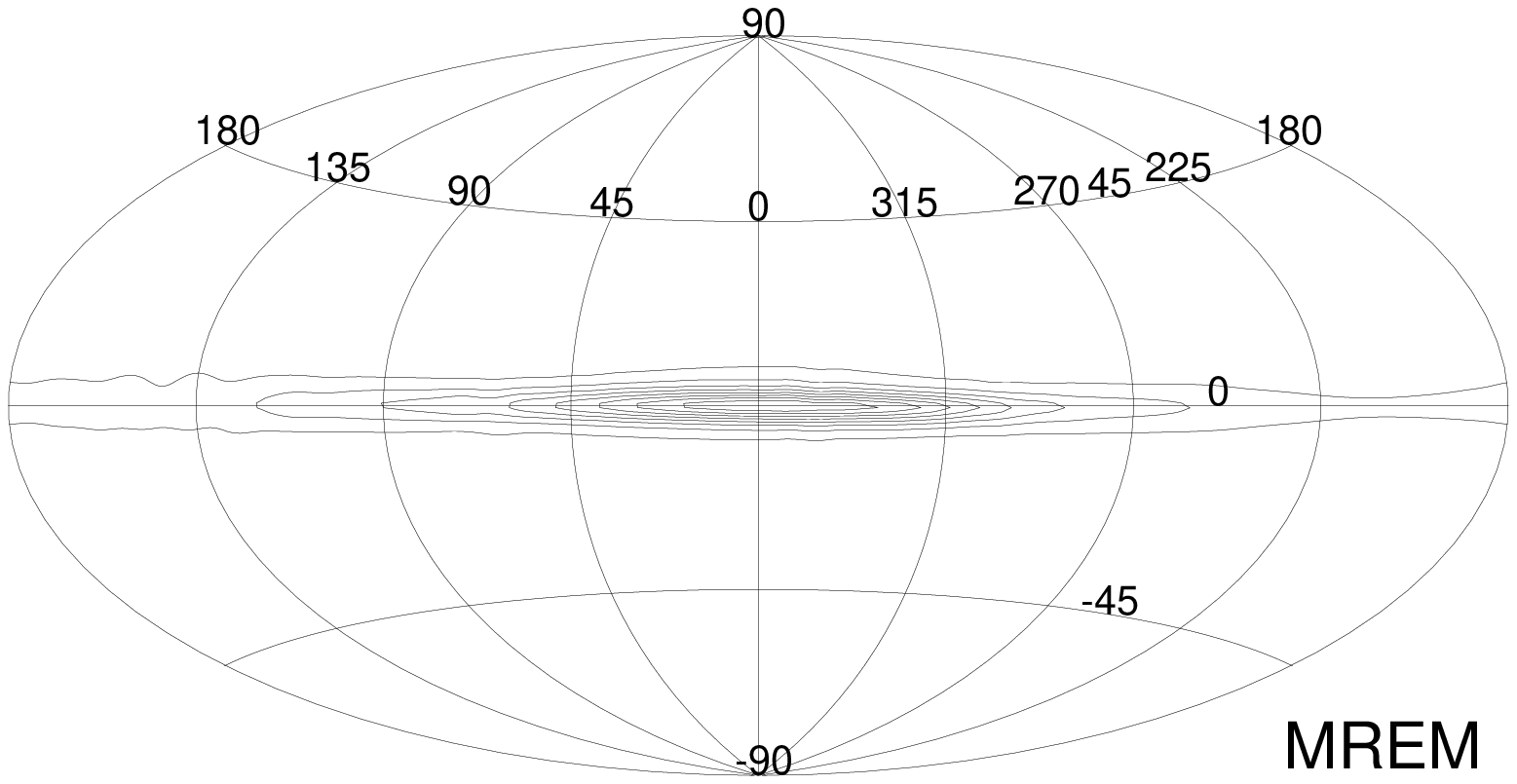, width=6.3cm, height=3cm}
            \hfill
            \psfig{file=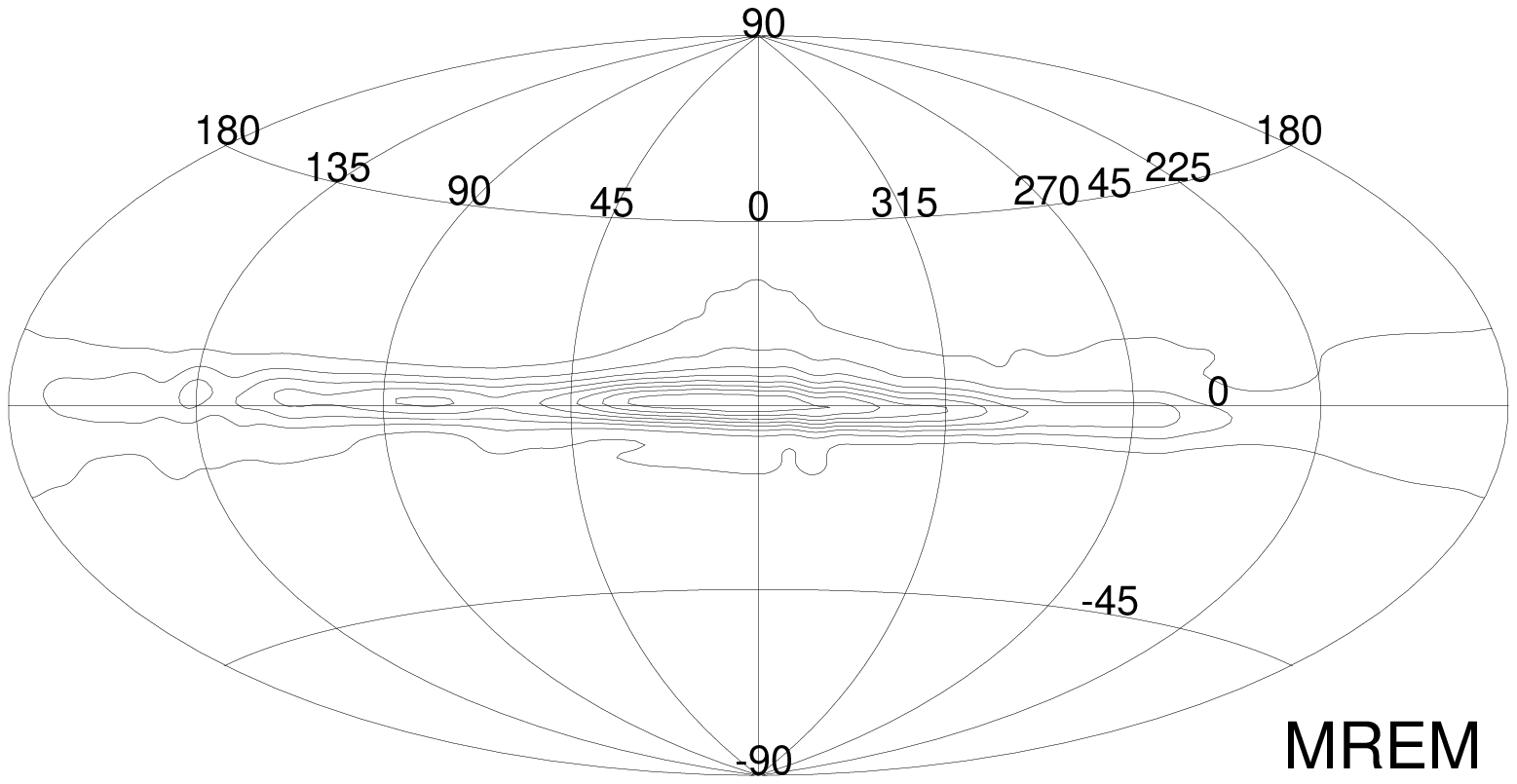, width=6.3cm, height=3cm}}
\centerline{\psfig{file=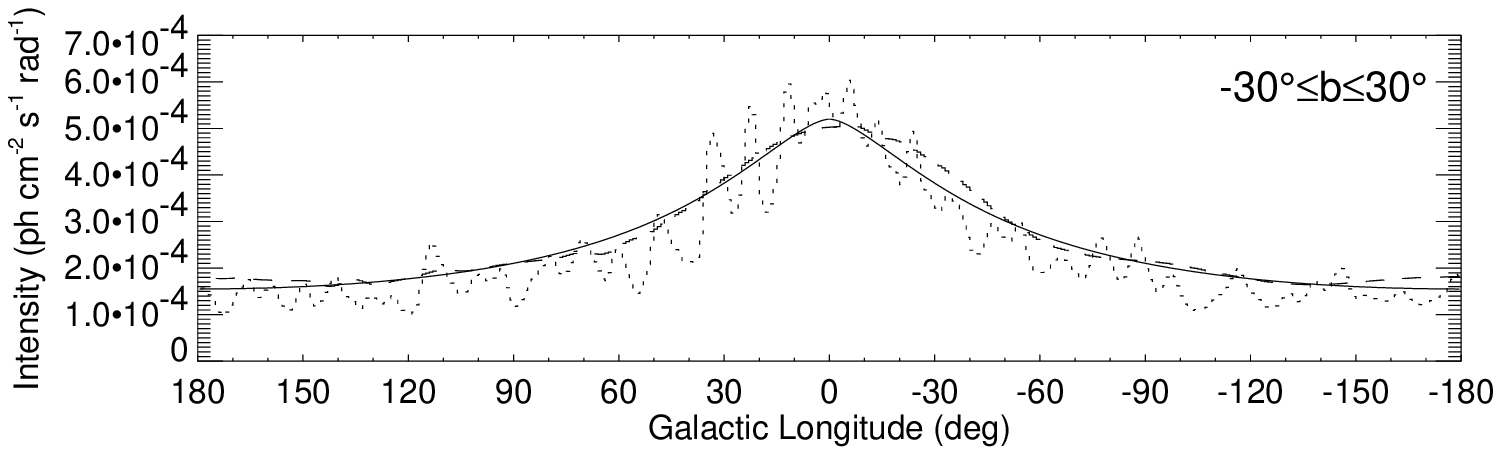, width=6.3cm}
            \hfill
            \psfig{file=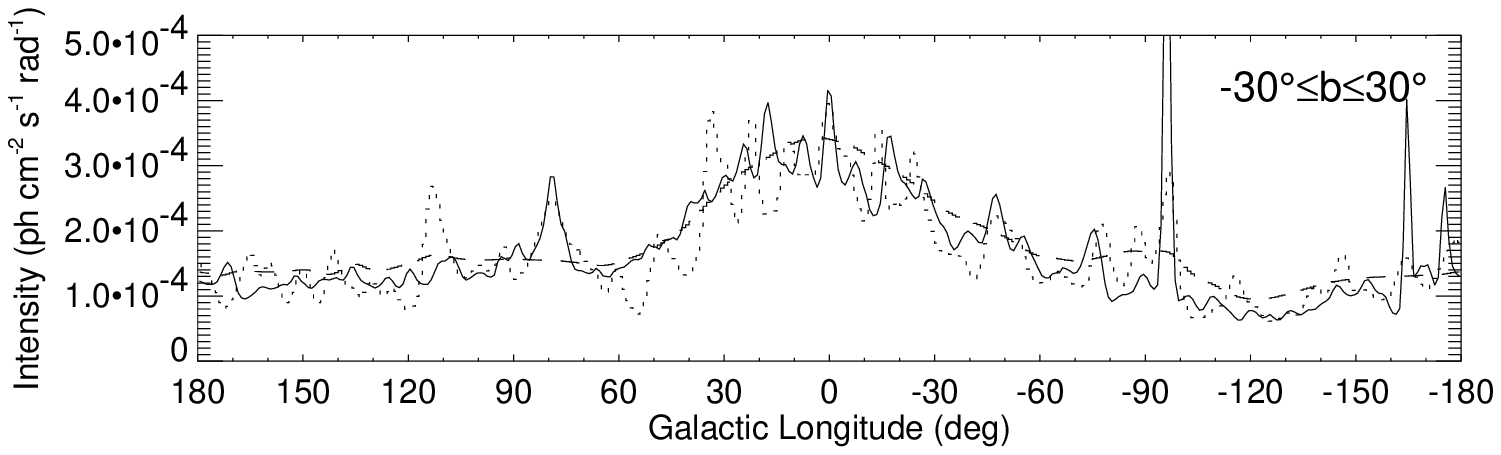, width=6.3cm}}
  \caption{FIGURE 1.
  Image reconstruction of two mock dataset assuming an exponential 
  disk ({\em left column}) and the EGRET $>100$ \MeV\ all-sky 
  map ({\em right column}) as source model.
  The bottom panels show the resulting longitude profiles 
  integrated over galactic latitudes of $|b|<30\deg$ ({\em solid:} 
  model, {\em dotted:} ME, {\em dashed:} MREM).
  } 
\end{figure}

The COMPTEL 1.8 \MeV\ all-sky images obtained by ME and MREM are 
presented in Fig.~2.
The ME reconstruction shows the typical lumpiness and hot spots which 
have also been seen for the exponential disk or EGRET map simulations.
We cannot decide which of the lumps may correspond to real emission 
and which are artefacts of the ME deconvolution, but comparison to the 
simulations suggests that many of them are not real.
The MREM image avoids this confusion, and shows a much smoother intensity 
profile with notable asymmetries with respect to the galactic center.
There is a sharp emission cut at $l=30\deg$ in contrast to a long emission 
tail out to $l=-120\deg$.
A prominent diffuse emission feature is seen in the Cygnus region, a 
weaker diffuse emission spots is located near the anticenter.
However, the excess emission in Carina ($l=-75\deg$) and Vela 
($l=-95\deg$), which is seen in the ME map, is almost imperceptible
in the MREM map.

\bsk
\ni 4. CONCLUSIONS
\ssk
\ni
\label{sec:conclusions}

An alternative COMPTEL 1.8 \MeV\ all-sky image is presented in this paper 
which has been derived using a multiresolution reconstruction algorithm 
based on wavelets.
Simulations suggest that our new algorithm provides a more reliable 
reconstruction of diffuse and smooth $\gamma$-ray emission than the maximum 
entropy algorithm which has been used before.
In particular, the number of `hot spots' and the lumpiness of the 
1.8 \MeV\ image is largely reduced with respect to the ME map, but 
some irregularities persist.
The simulations indicate that weak (3-4$\sigma$) point sources 
embedded in diffuse emission are not recovered by the algorithm, which 
explains the absence of prominent emission features in the MREM map 
towards Carina and Vela.
In this sense the algorithm is biased towards smoothness since it 
prefers explaining structures by smooth emission rather than point 
sources if the data are consistent with both.
MREM and ME are therefore complementary analysis tools, one clearly 
extracting the extended emission, the other providing hints for 
additional point-like emission.

\begin{figure}
\centerline{\psfig{file=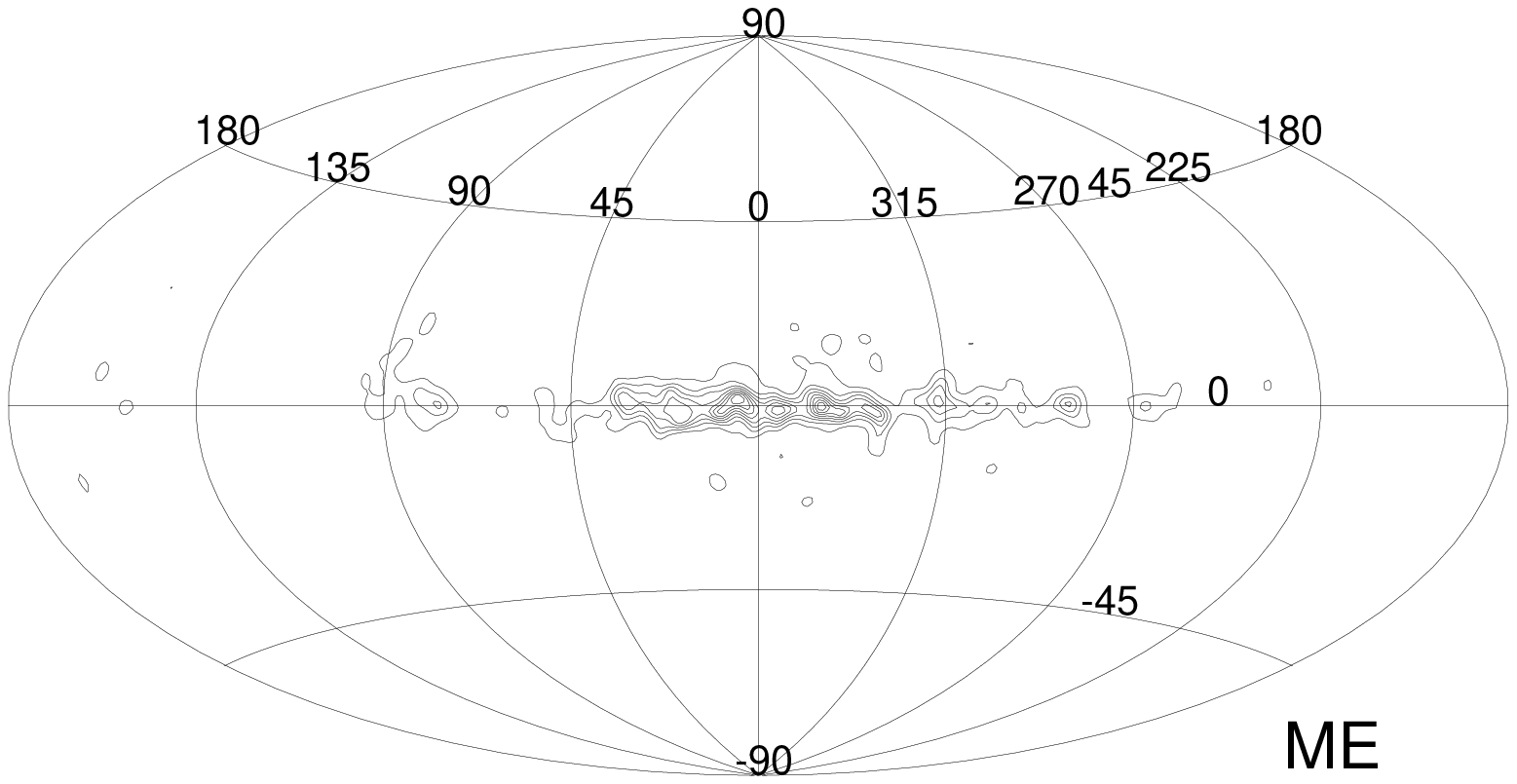, width=6.3cm}
            \hfill
            \psfig{file=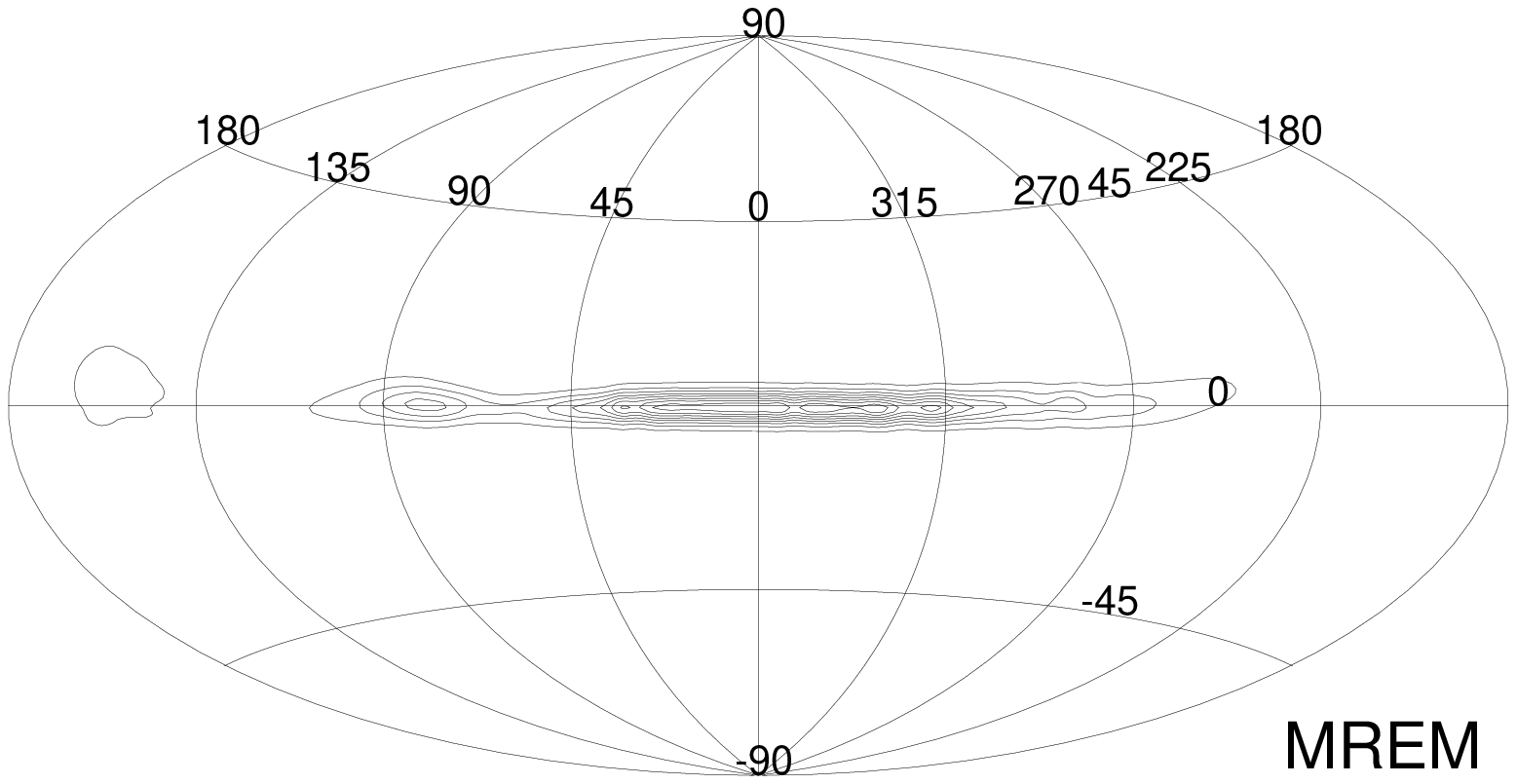, width=6.3cm}}
\centerline{\psfig{file=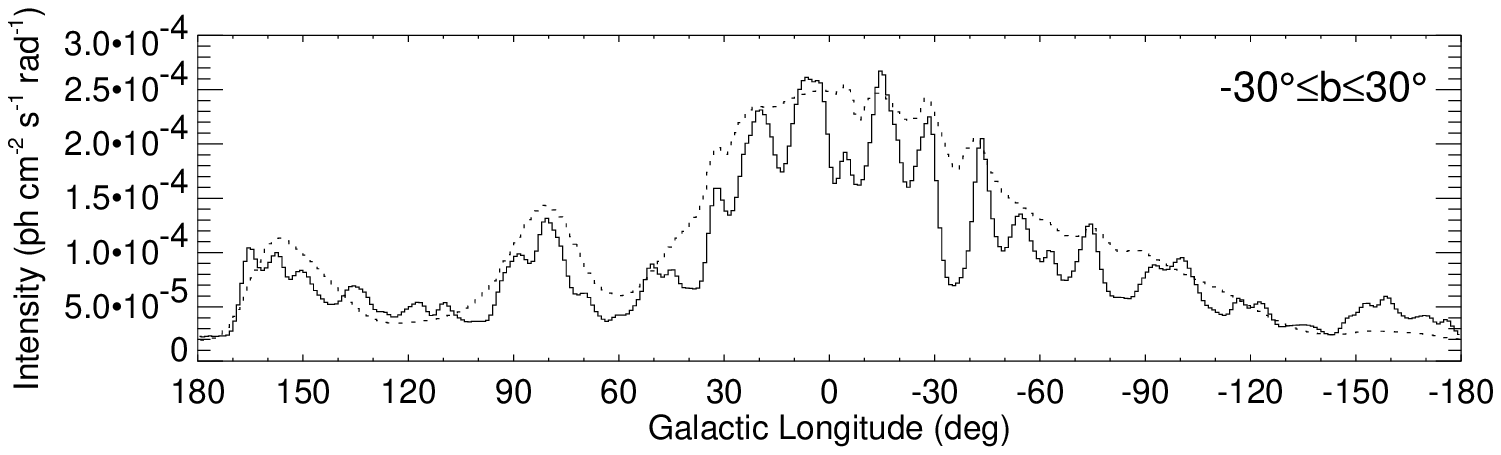, width=12.7cm}}
  \caption{FIGURE 2.
  COMPTEL 1.8 \MeV\ all-sky maps obtained by ME and MREM.
  The bottom plot compares the longitude profiles of both 
  reconstructions ({\em solid:} ME, {\em dotted:} MREM).
  } 
\end{figure}

\bsk
\baselineskip = 12pt

{\references \ni REFERENCES
\ssk
\ref     Dempster, A., Laird, N., Rubin, D.~1977, J.~Royal 
         Stat.~Soc.~Ser.~B, 39, 1
\ref     Diehl, R., et al.~1995, A\&A, 298, 445
\ref     Dixon, D., et al.~1998, in preparation
\ref     Kn\"odlseder, J., et al.~1996, \procspie, 2806, 386
\ref     Lucy, L. B.~1974, \aj, 79, 745
\ref     Oberlack, U., et al.~1996, \aaps, 120, C311
\ref     Richardson, W. H.~1972, J.~Opt.~Soc.~Am., 62, 55
\ref     Strong, A., et al.~1992, in DATA ANALYSIS IN ASTRONOMY - IV, 
         ed.~D.~Ges\`u, p. 251
         
}                      

\end{document}